\documentclass[a4paper]{jpconf}
\usepackage{graphicx}
\begin{document}
\title{ALICE results on quarkonium production in pp, p-Pb and Pb-Pb collisions}

\author{Giuseppe Eugenio Bruno for the ALICE Collaboration}

%\address{Production Editor, \jpcs, \iopp, Dirac House, Temple Back, Bristol BS1~6BE, UK}
\address{Dipartimento Interateneo di Fisica ``M. Merlin'' and Sezione INFN, Bari, Italy}
\ead{Giuseppe.Bruno@ba.infn.it}

\begin{abstract}
%The ALICE detector provides excellent capabilities to study quarkonium production 
%at the Large Hadron Collider (LHC) at both central and forward rapidity.  
The study of 
quarkonia, bound states of heavy (charm or bottom) quark-antiquark pairs such as the J/$\psi$ or the $\Upsilon$, %are expected to be produced by initial hard processes. 
%Thus they provide 
provides
insight into the earliest and hottest stages of %ultra-relativistic 
high-energy 
nucleus-nucleus collisions where the formation of a 
Quark-Gluon Plasma is expected. %At LHC energy, due to the abundant production of ${\rm c \bar{c}}$  pairs, 
%different production mechanisms has been suggested for the J/$\psi$ meson, where quark pairs (re-)combine in the 
%QGP or at the phase boundary, thus leading to an enhanced production of J/$\psi$ compared 
%to lower energy nuclear collisions and to pp collisions. 
High-precision data from proton-proton collisions represent an essential baseline for the  measurement of nuclear modifications in 
nucleus-nucleus collisions and serve also as a crucial test for models of quarkonium  hadroproduction. 
Another fundamental tool to understand the quarkonium production in nucleus-nucleus collisions is the the study of proton-nucleus 
interactions, 
which allows one 
to investigate cold nuclear matter effects, such as parton shadowing or gluon saturation. 
%At LHC energies, moreover, the density of final state particle in p-A collisions reaches values that can become comparable to semi-peripheral A-A collisions at lower energies. 
%Therefore final state dense matter effects in p-A collisions at LHC energies are not excluded. 
The ALICE detector provides excellent capabilities to study quarkonium production 
at the Large Hadron Collider at both central and forward rapidity.  
An overview on ALICE results on quarkonium production in pp, p--Pb and Pb--Pb collisions is presented. Results are compared to theoretical model predictions. 
\end{abstract}

\section{Introduction}
The ALICE experiment~\cite{ALICE} studies proton-proton (pp), proton-nucleus (p--A) and nucleus-nucleus (A--A) 
collisions at the Large Hadron Collider (LHC), with the main goal of investigating the properties of the 
high-density, colour-deconfined state of strongly-interacting matter (the Quark-Gluon Plasma, QGP) 
that is expected to be formed in high-energy nuclear collisions. 
%The formation of a QGP %in high-energy nuclear collisions 
%can be evidenced in a variety of ways. 
%One of its most striking expected
%signatures is 
The suppression of quarkonium
states was proposed long ago~\cite{Matsui:1986dk} 
as a signature of the formation of the QGP: 
%both of the charmonium  and the bottomonium families. 
%This is thought to be a direct effect of
%deconfinement, when 
in a deconfined medium, 
the binding potential between the constituents of
a quarkonium state, a heavy quark (Q) and its antiquark (${\rm \bar{Q}}$), is screened by
the colour charges of the surrounding light quarks and gluons.  %The
%suppression is predicted to occur above the critical temperature of
%the medium %($T_c$) 
%and depends on the ${\rm Q\bar{Q}}$ binding energy. 
In this scenario, quarkonium suppression occurs sequentially, according to the binding
energy of each meson: strongly bound states, as the ${\rm J/\psi}$ and ${\rm \Upsilon(1S)}$, 
should melt at higher temperatures with respect to the more loosely bound ones, as the ${\rm \psi(2S)}$ and ${\rm \chi_c}$, 
for the charmonium family, or the ${\rm \Upsilon(2S)}$ and ${\rm \Upsilon(3S)}$ for the bottomonium one.
%temperature. 
%Examples of dissociation temperatures are given in
%Ref.~\cite{Mocsy:2007jz}: $T_{\rm{dissoc}}\sim\!1\,T_c$, $1.2\,T_c,$
%and $2\,T_c$ for the ${\rm \Upsilon(3S)}$, ${\rm \Upsilon(2S)}$, and ${\rm \Upsilon(1S)}$, respectively. Similarly,
%in the charmonium family the dissociation temperatures are
%$\leq1\,T_c$ and $1.2\,T_c$ for the $\psi'$ and \Jpsi,
%respectively. 
The in-medium modification of quarkonium production is usually quantified by the nuclear modification factor: 
$R_{\rm AA} = \frac{{\rm d^2}N_{\rm AA}/{\rm d} p_{\rm T} {\rm d} y }{\left<  N_{\rm coll} \right> \, 
                    {\rm d^2}N_{\rm pp}/{\rm d} p_{\rm T} {\rm d} y }$, where 
${\rm d^2}N/{\rm d} p_{\rm T} {\rm d} y $
denotes the transverse momentum ($p_{\rm T}$) and rapidity ($y$) differential yield of a given particle measured 
in A--A or pp collisions and $\left<  N_{\rm coll} \right>$ is the average number of nucleon-nucleon collisions over 
the given centrality interval of A--A collisions; $\left<  N_{\rm coll} \right>$ is calculated using the 
Glauber model~\cite{Miller}.  
However, there are further possible effects on the
quarkonium production in heavy-ion collisions. On the one hand,
the large
number of heavy quarks  produced in heavy-ion collisions, in particular in the charm sector 
at the energies accessible by the LHC, may
lead to an increased production of charmonia via 
 (re-)combination in the QGP~\cite{reco1,reco2} or at the phase boundary~\cite{reco3,reco4}.   
%statistical recombination~\cite{Zhao:2010nk,Andronic:2006ky,Capella:2007jv,Thews:2005vj,Yan:2006ve,Grandchamp:2005yw}.
On the other hand, 
modifications to the parton distribution functions inside the nucleus
(shadowing) and other cold nuclear matter effects can change the
production of quarkonia without the presence of a
QGP~\cite{reco3, Vogt:2010aa}. Hence, for the interpretation of A--A results, data from p--A 
collisions are crucial because they allow us to disentangle these cold nuclear matter effects from those related to 
the formation of a hot QCD medium.  
Finally, besides being the natural reference for A--A studies, quarkonium hadro-production in 
 elementary pp collisions are important {\it per se},  
representing a challenging testing ground for models based on QCD. 
ALICE results on quarkonium production in pp are therefore also summarized in the following. 
\section{Experiment,  data taking conditions and analysis}
The ALICE experiment consists of a central barrel embedded in a solenoidal magnet and a forward muon spectrometer. 
Details on the experimental set-up can be found in~\cite{ALICE}. The measurement of quarkonium production is carried out in the
central barrel ($|y| < 0.9$) through their ${\rm e^+e^-}$ decay, while at forward rapidity ($2.5 < y < 4$) the ${\rm \mu^+\mu^-}$  decay is studied 
in the muon spectrometer.  

For the measurement in the central barrel described hereafter, events are triggered by a minimum bias (MB) condition, which corresponds %basically  
roughly 
to the presence of at least one charged particle in about 8 units of pseudo-rapidity. This condition is defined  using the 
information from the two innermost pixel layers of the silicon Inner Tracking System 
(ITS) ($|\eta|< 2$ and $|\eta|<1.4$) and from two arrays of scintillators (VZERO) placed 
at forward rapidities ($-3.7< \eta <-1.7$ and $2.8 < \eta < 5.1$). 
In Pb-Pb, several centrality classes can be further defined at the trigger level by means of thresholds on the total signal amplitude in the VZERO. 
For the measurement with the forward muon spectrometer one can require, in addition to the MB condition, one or two %(depending on the data taking) 
candidate muons to be detected by the muon triggering system.
 
For Pb--Pb collisions at central rapidity, the analysis is based on a combination of 2010 data, taken with a MB trigger, and 2011 data where, 
in addition, centrality selections corresponding to  central events (0--10\%) and semi-central events (10--40\%)
were performed. The integrated luminosity is $L_{\rm int}=15$~${\rm \mu b^{-1}}$. The Pb--Pb forward rapidity results
refer to the 2011 data ($L_{\rm int}=70$~${\rm \mu b^{-1}}$), and were collected by requiring a dimuon trigger, which can
select both opposite- and like-sign pairs. 
In order to define $R_{\rm AA}$, one needs a sample of pp data collected at the same energy ($\sqrt{s}=2.76$~TeV) and in the same kinematic domain of
Pb-Pb data. Such data were taken in 2011, collecting $L_{\rm int}=1.1$~${\rm nb^{-1}}$ with a MB trigger for the
central rapidity analysis and $L_{\rm int}=20$~${\rm nb^{-1}}$ with a single-muon trigger for the forward rapidity
analysis. Moreover, pp data at $\sqrt{s}= 7$~TeV, the top energy of the 2010 and 2011 LHC data taking, were
also collected and analysed, the results %presented 
discussed  
here referring to integrated luminosities up to
100~${\rm nb^{-1}}$. 

Data were collected in 2013 for p-Pb collisions using, for the quarkonium studies with the muon spectrometer, 
the % opposite-sign 
dimuon trigger in coincidence with 
the MB trigger. Due to the energy asymmetry of the LHC beams ($E_{\rm p} = 4 $~TeV, $E_{\rm Pb} = 1.58 \, {\rm A}$~TeV) 
the nucleon-nucleon
center-of-mass system of the collisions does not coincide with the laboratory system, but is shifted by
$\Delta y = 0.465$ in the direction of the proton beam. Data have been taken with two beam configurations, by
inverting the direction of the orbits of the two particle species. 
In this way in the dimuon channel the regions $2.03 <y_{\rm cms} < 3.53$
%rapidity evaluated in the nucleon-nucleon centre of mass frame
and $-4.46 < y_{\rm cms} < -2.96$ have been studied, where positive rapidities refer to the situation where
the proton beam is travelling towards the muon spectrometer (in the following these configurations are
referred to as p-Pb and Pb-p, respectively). The integrated luminosities of the analysed data for the two
configurations are about 5~${\rm nb^{-1}}$ (p-Pb) and 6~${\rm nb^{-1}}$ (Pb-p). 
For the central barrel studies, about 130 M MB events (i.e. an integrated luminosity of about 50~$\mu b^{-1}$) have been collected.  In addition, other  rarer trigger data sample, of interest for charmonia studies, 
e.g. that requiring an electron in the Transition Radiation Detector (TRD), corresponding to $L_{\rm int} \approx 1.5$~${\rm nb^{-1}}$ are also available. These data samples are being analysed at the time of the conference. 

\begin{figure}[hbt]
\resizebox{0.35\textwidth}{!}{
\includegraphics{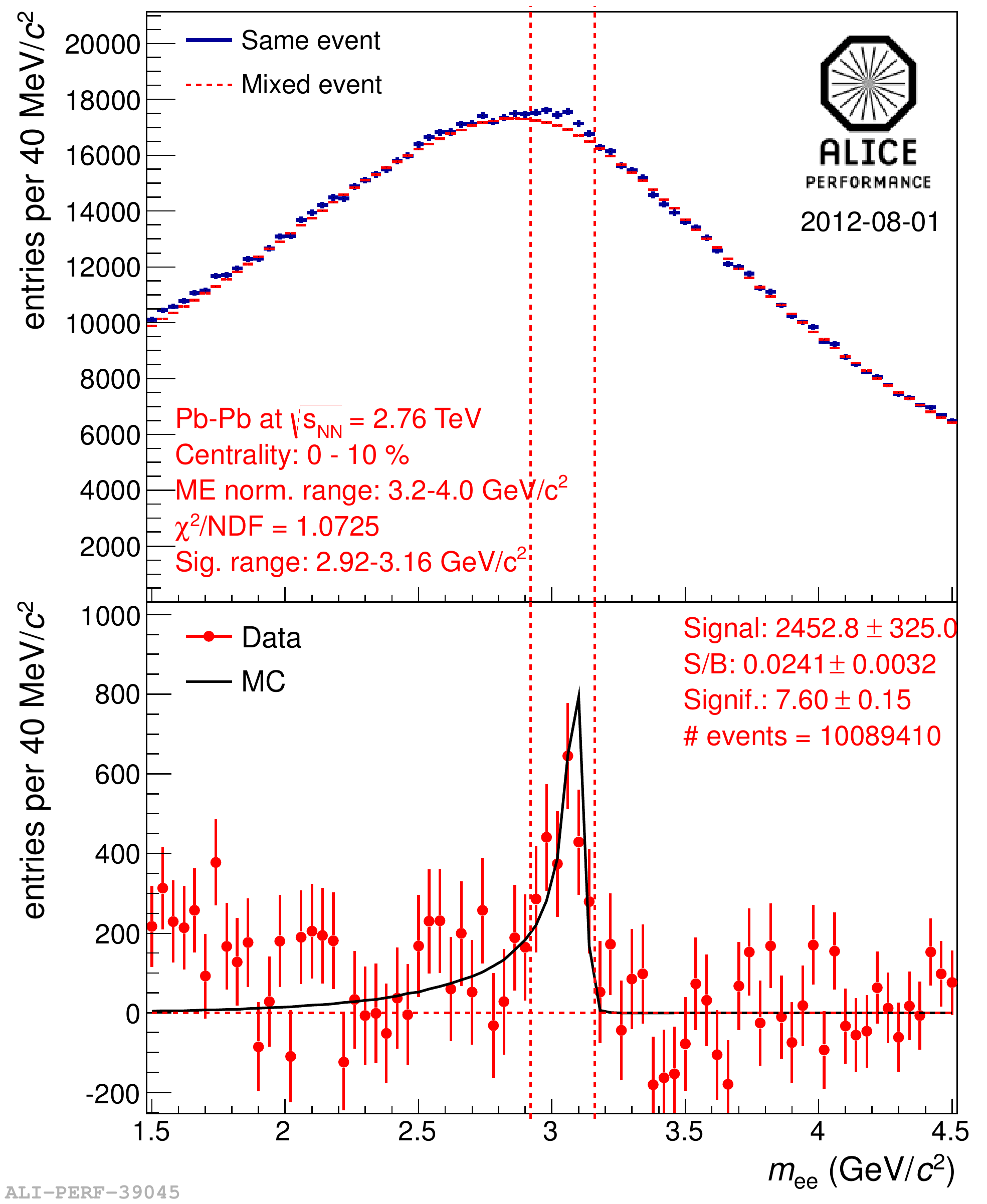}}
\resizebox{0.35\textwidth}{!}{
\includegraphics{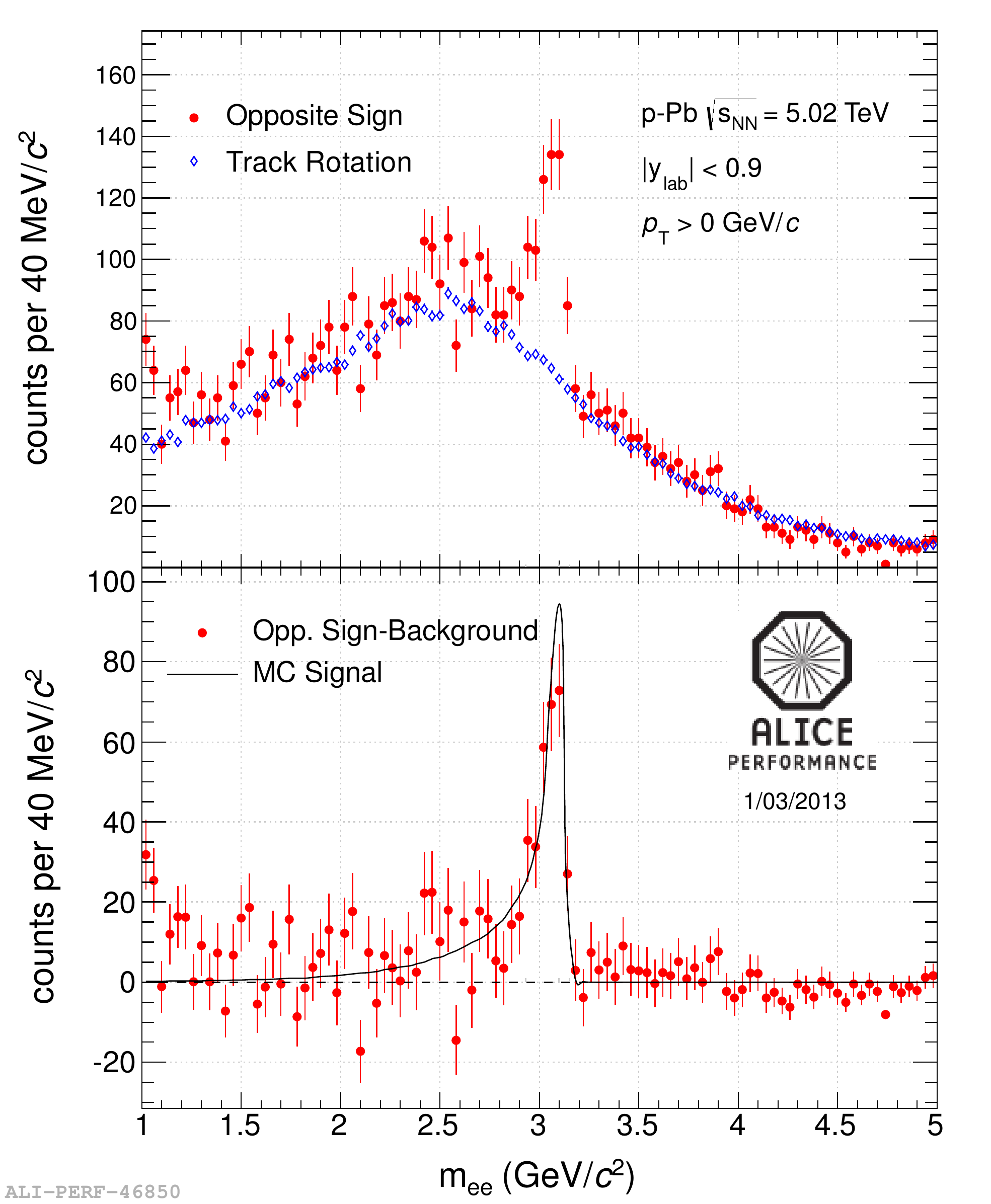}} \\
\resizebox{0.32\textwidth}{!}{
\includegraphics{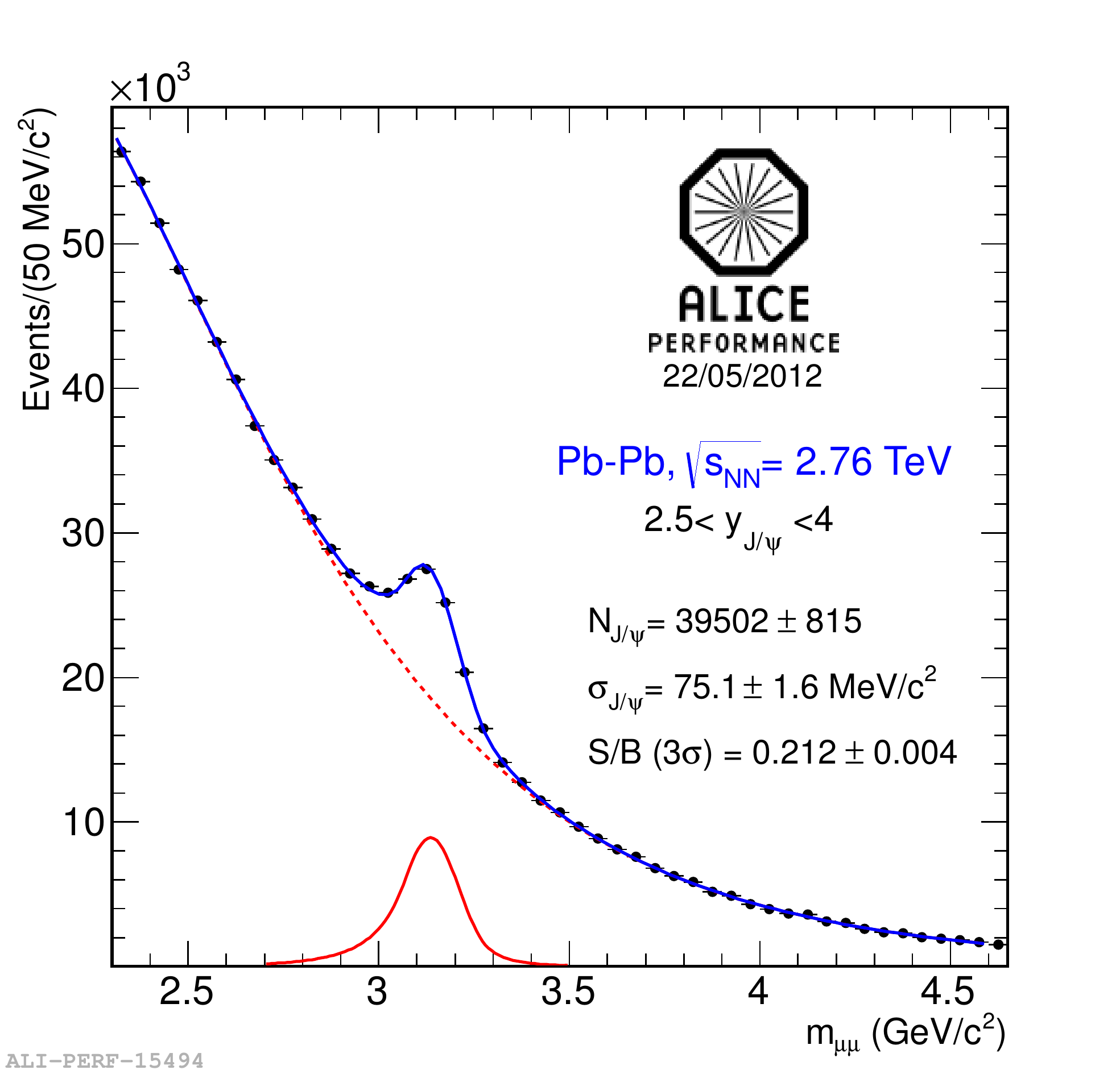}}
\resizebox{0.37\textwidth}{!}{
\includegraphics{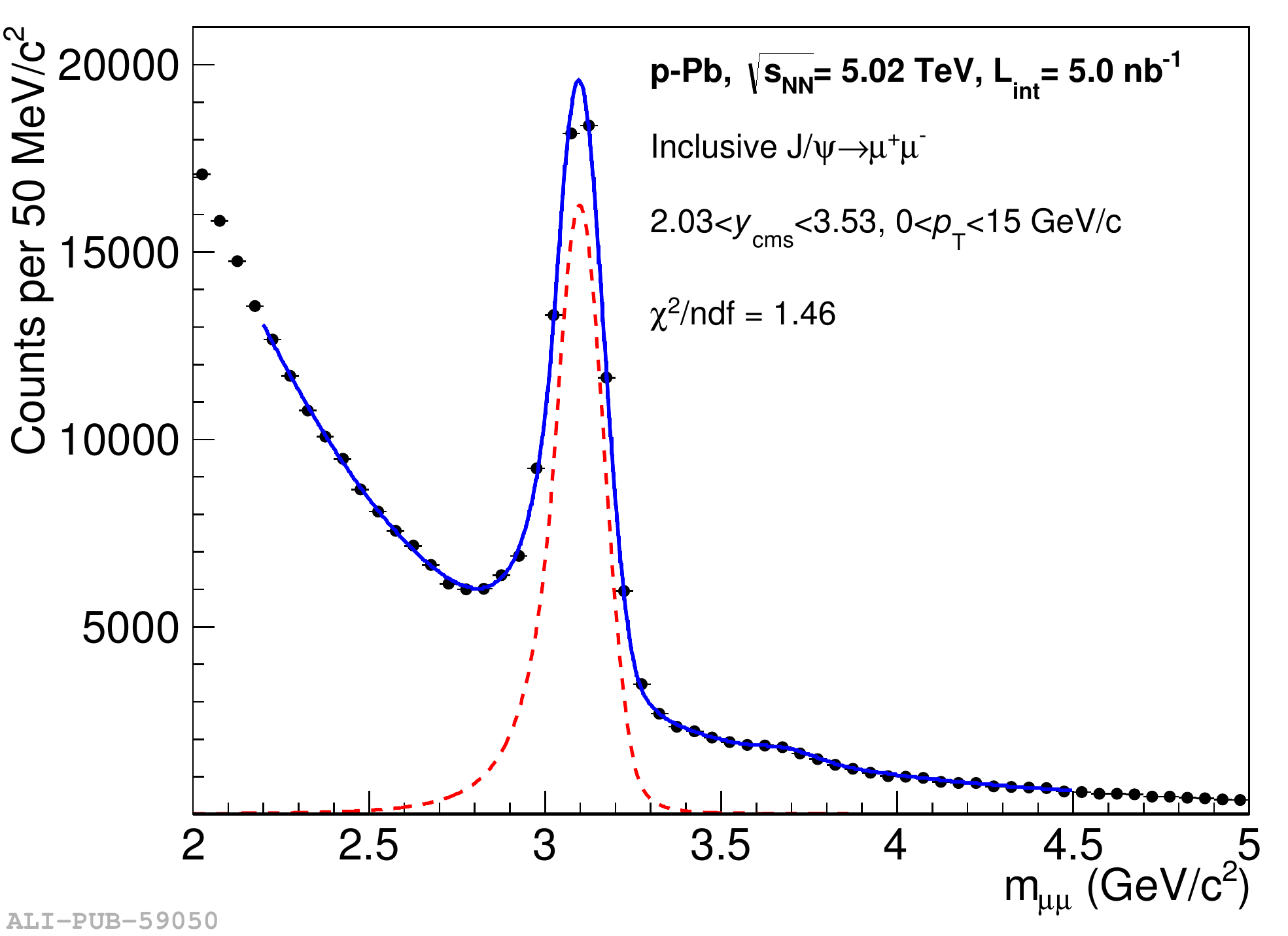}}\\
\resizebox{0.35\textwidth}{!}{
\includegraphics{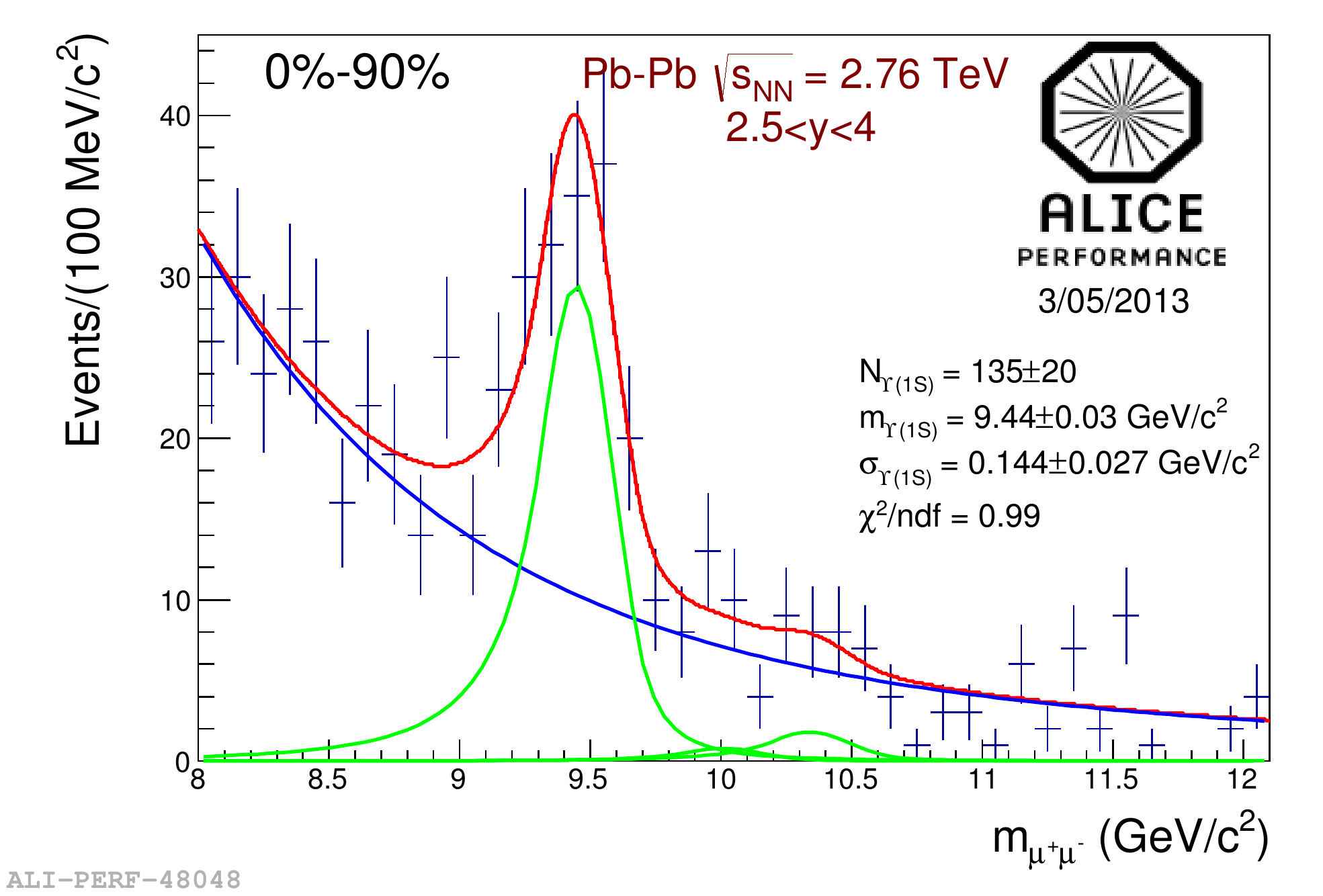}}
\resizebox{0.35\textwidth}{!}{
\includegraphics{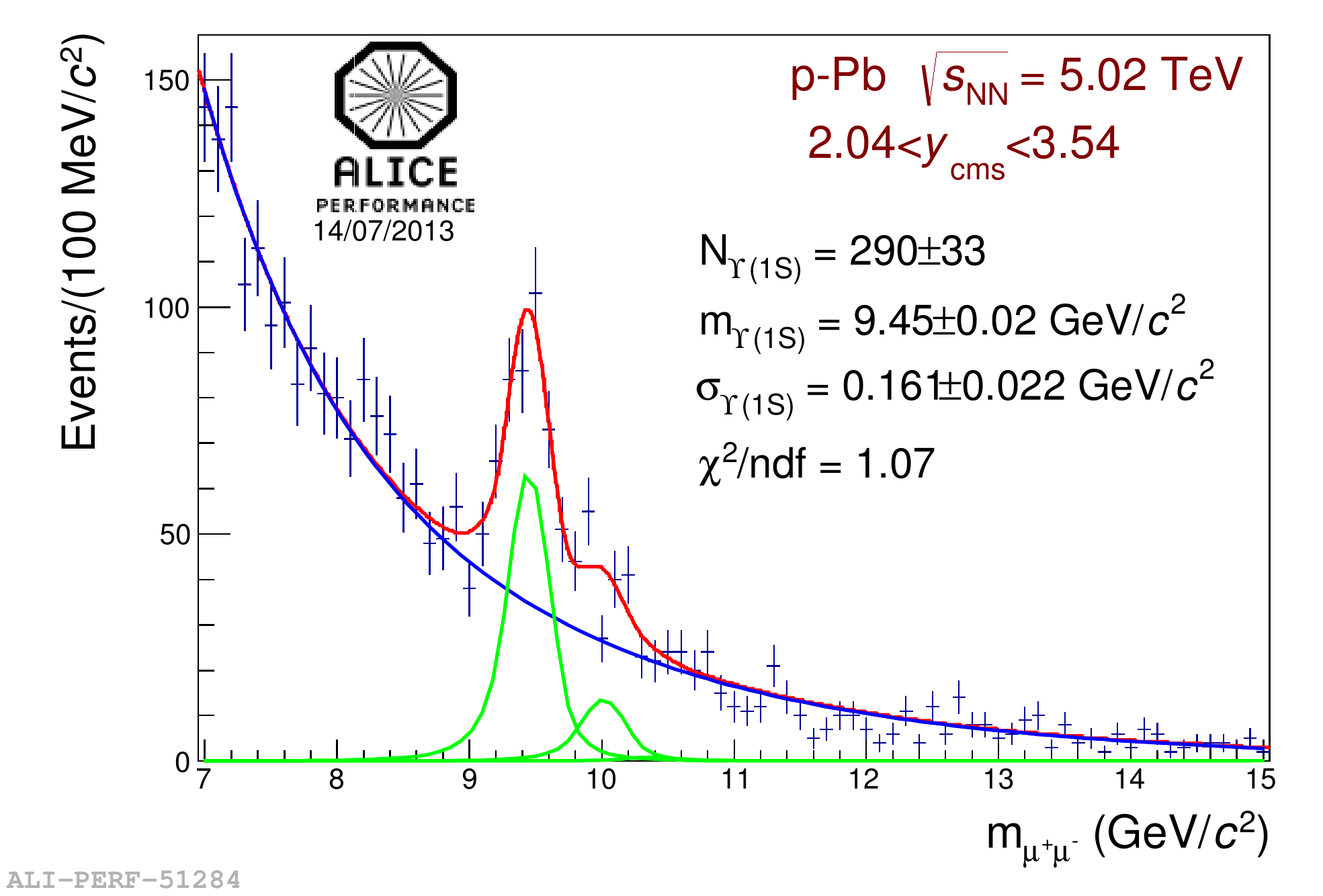}}
\caption{\label{fig:Mass}Invariant mass distributions of opposite sign electron (top panels) 
                         or muon (middle and bottom panels) pairs in Pb--Pb (left panels) 
                         and p--Pb (right panels) collisions. In the dielectron channel 
                         the combinatorial background is described with the event mixing 
			 or the track rotation methods and subtracted from the same-event opposite-sign
			 distributions.  
       			  }
\end{figure}

In the central barrel the analyses are mainly based on the charged track reconstruciton performed with 
the ITS and Time Projection Chamber (TPC) detectors, on the electron identification via the measurement of the specific energy 
loss in the TPC and on the auxilary information from the Time Of Flight (TOF) detector and the TRD for the rejection of hadrons. 
Thanks to the excellent spatial resolution of the ITS, the non-prompt J/$\psi$ contribution can be separated, 
thus allowing a measurement of beauty production down to very low $p_{\rm T}$ ($p_{\rm T}({\rm J/\psi}) 
\approx 1.5$~GeV/$c$). Details of the analyses are discussed in~\cite{Fiorella}.   
In the muon channel, opposite sign (OS) muons, which have been 
filtered by the hadron absorber and reconstructed in the muon spectrometer, are paired to obtain 
invariant mass distributions with clean mass peaks for the quarkonium states. 
Quarkonium signals are extracted by means of fit to the OS mass 
distribution, using a phenomenological shape for the background, and Crystal Ball shape for the signals.
Further details can be found in~\cite{Lizardo,Palash,Igor}.
Examples of the dileptons invariant mass distributions 
obtained in Pb--Pb and p--Pb collisions are shown in Fig.~\ref{fig:Mass}.

\section{Results}
\subsection{pp}
ALICE has measured the production cross section of inclusive J/$\psi$ at $\sqrt{s}=7$ and 2.76~TeV~\cite{sedici,diciassette}. 
At $\sqrt{s}=2.76$~TeV a $p_{\rm T}$ differential study was possible only in the forward region, due to the rather small MB integrated 
luminosity. At central rapidity, the fraction of J/$\psi$ from beauty hadron decay could be separated at low $p_{\rm T}$~\cite{ventuno}, which also led to 
an estimation of the total ${\rm b \bar{b}}$ cross section at $\sqrt{s}=7$~TeV. 
The first LHC study of the inclusive J/$\psi$ polarization has been performed at forward rapidity, showing that the J/$\psi$  production is essentially 
unpolarized up to $p_{\rm T} = 8$~GeV/$c$~\cite{diciotto}. The comparisons of these results with NLO NRQCD calculations that include color octet processes 
show good agreement. We have also performed a unique  measurement of the J/$\psi$  yield as a function of the charged 
particle multiplicity~\cite{multi,renu}, which shows an almost linear increase of the yield with the multiplicity. The latter result 
may either indicate that J/$\psi$ production in pp is connected with 
strong hadronic activity, or that multiparton interactions could also affect the harder momentum scales relevant for quarkonium production.    

Preliminary results on $\Upsilon$ %(and ${\rm \Upsilon(2S)}$) 
production in  pp collisions at $\sqrt{s}=7$ TeV have also been delivered~\cite{Palash}, 
which are in good agreement with the published LHCb results~\cite{LHCbUpsilonpp}.  

Reference pp data at the same energies %(in the nucleon-nucleon center of mass frame) 
of the p--Pb ($\sqrt{s_{\rm NN}}=5.02$~TeV) and Pb--Pb ($\sqrt{s_{\rm NN}}=2.76$~TeV) 
interactions are, respectively, not available and of limited statistics 
(a few days of data taking was devoted to the study of pp collisions at $\sqrt{s}=2.76$~TeV). 
Therefore, interpolation procedures are often introduced (except for the forward rapidity J/$\psi$   
results at $\sqrt{s}=2.76$~ TeV) 
to estimate the production cross section of quarkonium states at these energies,  
which lead to large uncertainties.  

\subsection{Pb--Pb}
\begin{figure}[h]
\resizebox{0.33\textwidth}{!}{
\includegraphics{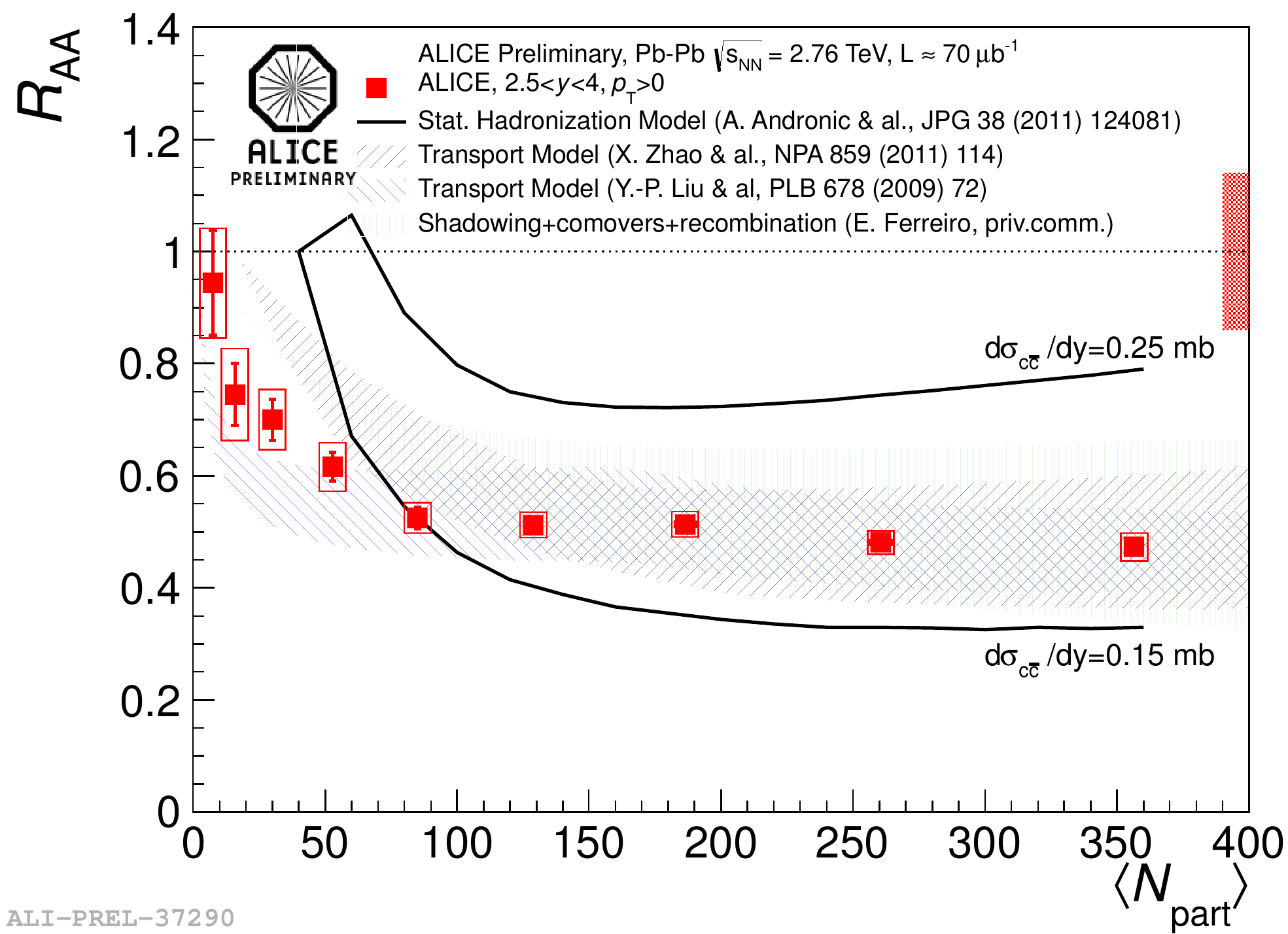}}
\resizebox{0.33\textwidth}{!}{
\includegraphics{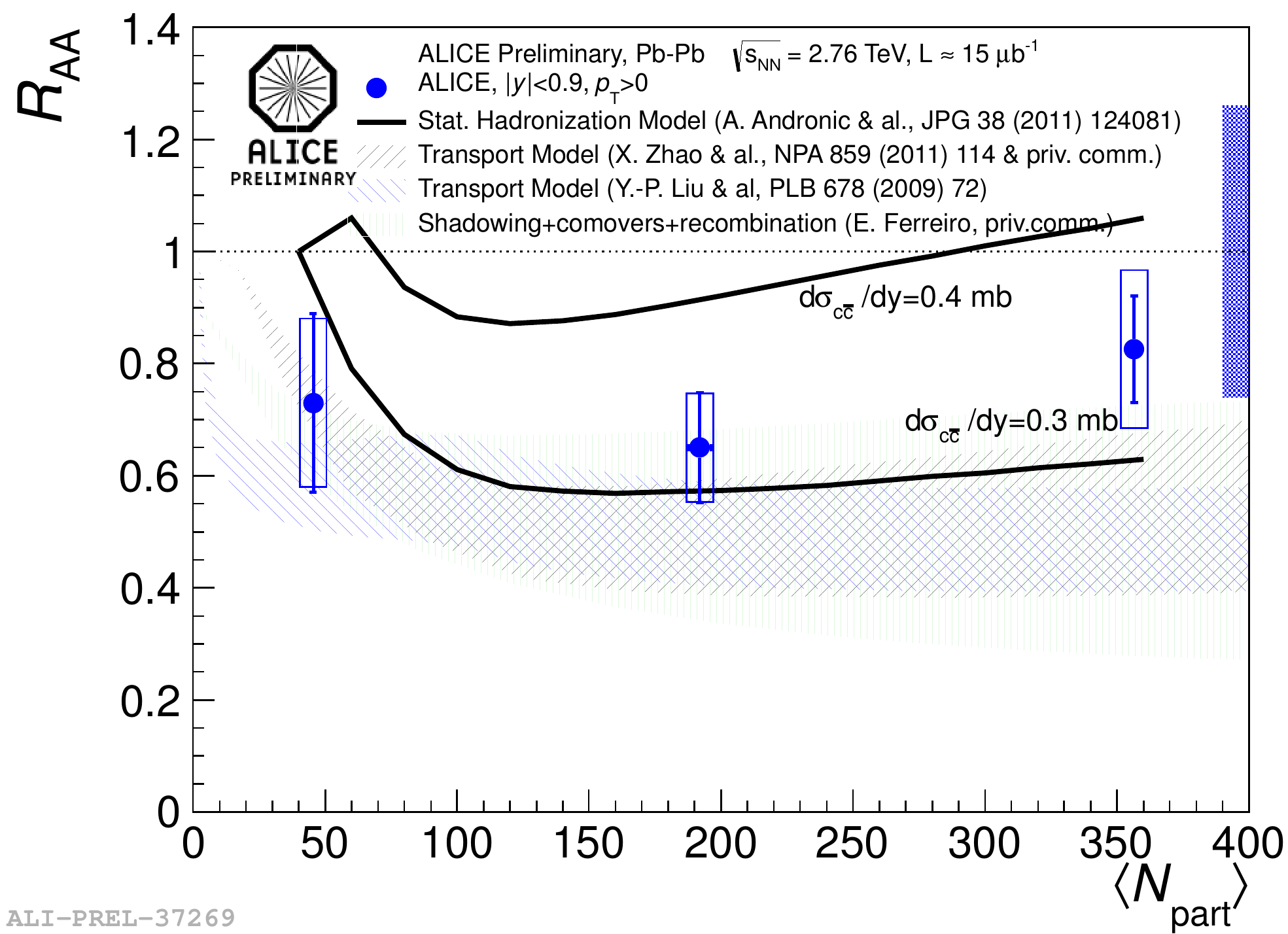}}
\resizebox{0.33\textwidth}{!}{
\includegraphics{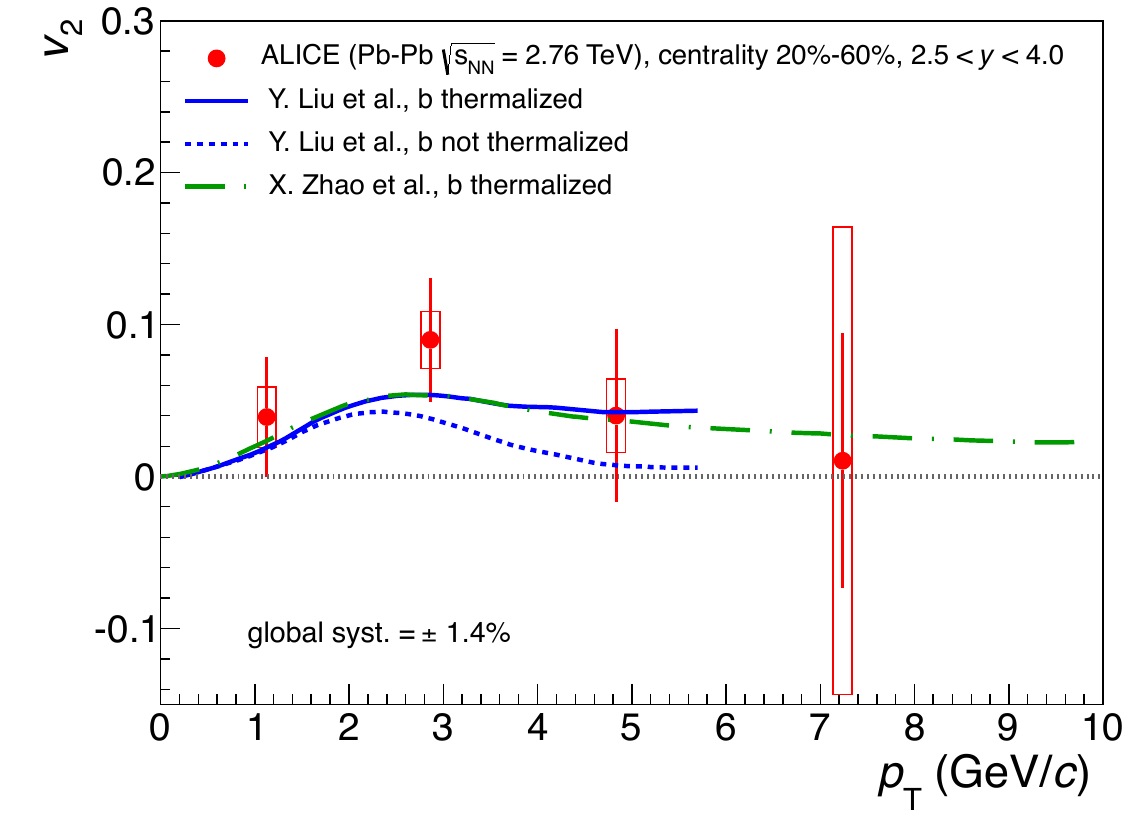}}
\caption{\label{Raa}
Inclusive J/$\psi$ $R_{\rm AA}$ measured in Pb--Pb collisions at $\sqrt{s_{\rm NN}} = 2.76$ TeV at forward (left panel) and central (middle pannel)
rapidity %compared to theoretical models (see~\cite{Enrico} and reference therein). %The point to point uncorrelated systematic uncertainties are represented as boxes around the data points, while the 
%statistical ones are shown as vertical bars. %Global correlated systematic uncertainties are quoted directly in the legend.
and inclusive J/$\psi$ $v_2(p_{\rm T})$ for semi-central %(20--60\%) 
Pb-Pb collisions (right panel) %$\sqrt{s_{\rm NN}} = 2.76$ TeV, 
compared to theoretical models. % (see text for details).  
}
\end{figure}

%The centrality dependence of 
% nuclear modification factor is compared with results obtained by PHENIX in Fig.~\ref{Raa}.
The inclusive J/$\psi$ nuclear modification factor as a function of
centrality, $p_{\rm T}$ and $y$ in Pb-Pb collisions at $\sqrt{s_{\rm NN}} = 2.76$~TeV, has been measured down to zero $p_{\rm T}$~\cite{Fiorella}. 
At forward rapidity, $R_{\rm AA}$ shows a clear suppression of the J/$\psi$ yield, with no significant dependence on 
centrality for $\left< N_{\rm part} \right>$ 
larger than 70 (left panel of Fig.~\ref{Raa}). At mid rapidity the J/$\psi$ $R_{\rm AA}$ is compatible with a 
constant value (middle panel of Fig.~\ref{Raa}). 
At forward rapidity
the J/$\psi$ $R_{\rm AA}$ exhibits a strong $p_{\rm T}$ dependence and decreases by a factor of 2 from low $p_{\rm T}$ to 
high $p_{\rm T}$. 
This behavior shows a striking difference with the one observed by the PHENIX experiment at RHIC energy~\cite{Phenix}, which 
measured a larger suppression at low $p_{\rm T}$ than that measured by ALICE at the LHC. 
The measurement of the inclusive $J/\psi$ elliptic flow at forward rapidity~\cite{elliptic} has provided an indication of non-zero $v_2$ in 
semi-central Pb--Pb collisions.  
Both results are in agreement with the global picture in which a significant fraction of the observed J/$\psi$ is 
produced from (re-)combination of charm quarks in the QGP phase or at the phase boundary.

%Average pt
The fraction of non-prompt J/$\psi$ ($f_{\rm B}$) has been measured at central rapidity in the range $2<p_{\rm T}< 10$~ GeV/$c$~\cite{Fiorella}. 
No significant dependence of $f_{\rm B}$ on centrality could be determined.  
Considering the ALICE and CMS results~\cite{CMS} together, an indication for a similar trend of $f_{\rm B}$ as a function of $p_{\rm T}$ in
pp and Pb--Pb collisions is observed. 
%Psi/J/psi
The preliminary results on the $\psi({\rm 2S})$ are discussed in~\cite{Enrico}.  %, are omitted here for reason of space. 
The nuclear modification factor of the $\Upsilon({\rm 1S})$ has been measured at forward rapidity 
($2.5 < y < 4.0$) for  $p_{\rm T}>0$~\cite{Palash}. Within the uncertainty a similar suppression as that measured for 
inclusive J/$\psi$ is observed. The suppression is stronger for central than semi-peripheral collisions. 
Combining this measurement with that of CMS~\cite{CMS2} performed for  $|y| < 2.4$,  
no rapidity dependence of the $\Upsilon({\rm 1S})$ suppression is  observed within the 
large rapidity range probed by the two experiments.  

\subsection{p--Pb}
%To fully understand the quarkonia production in nucleus-nucleus production 
The nuclear modification factor in p-Pb collisions, $R_{\rm pPb}$, has been measured in the muon spectrometer 
for the J/$\psi$ and $\Upsilon({\rm 1S})$ states~\cite{Palash,Igor}. %(see left panel of Fig.~\ref{Fig:R2}).  
Since no pp collisions at $\sqrt{s}=5.02$ TeV have been delivered, the pp references are obtained interpolating the 
results collected at higher and lower energies, and these pp references introduce an important contribution to the total 
uncertainty on $R_{\rm pPb}$, especially for the J/$\psi$. 
%Definition of R\_FB: pro and contros
The variable $R_{\rm FB}$, which is defined as the ratio of the J/$\psi$ (and similarly for the $\Upsilon$) forward to 
backward yields measured in the common  J/$\psi$ rapidity range %(in the nucleon-nucleon centre of mass frame) 
$2.96< |y_{\rm cms}| <3.53$, has been introduced to avoid the 
need of the pp reference, and its relative large uncertainty. 
The drawback of this observable, which is measured in a restricted rapidity interval,  
is  a minor sensitivity to discriminate among models that can describe the cold nuclear matter effects, with respect to $R_{\rm pPb}$.     

\begin{figure}[h]
\resizebox{0.45\textwidth}{!}{
\includegraphics{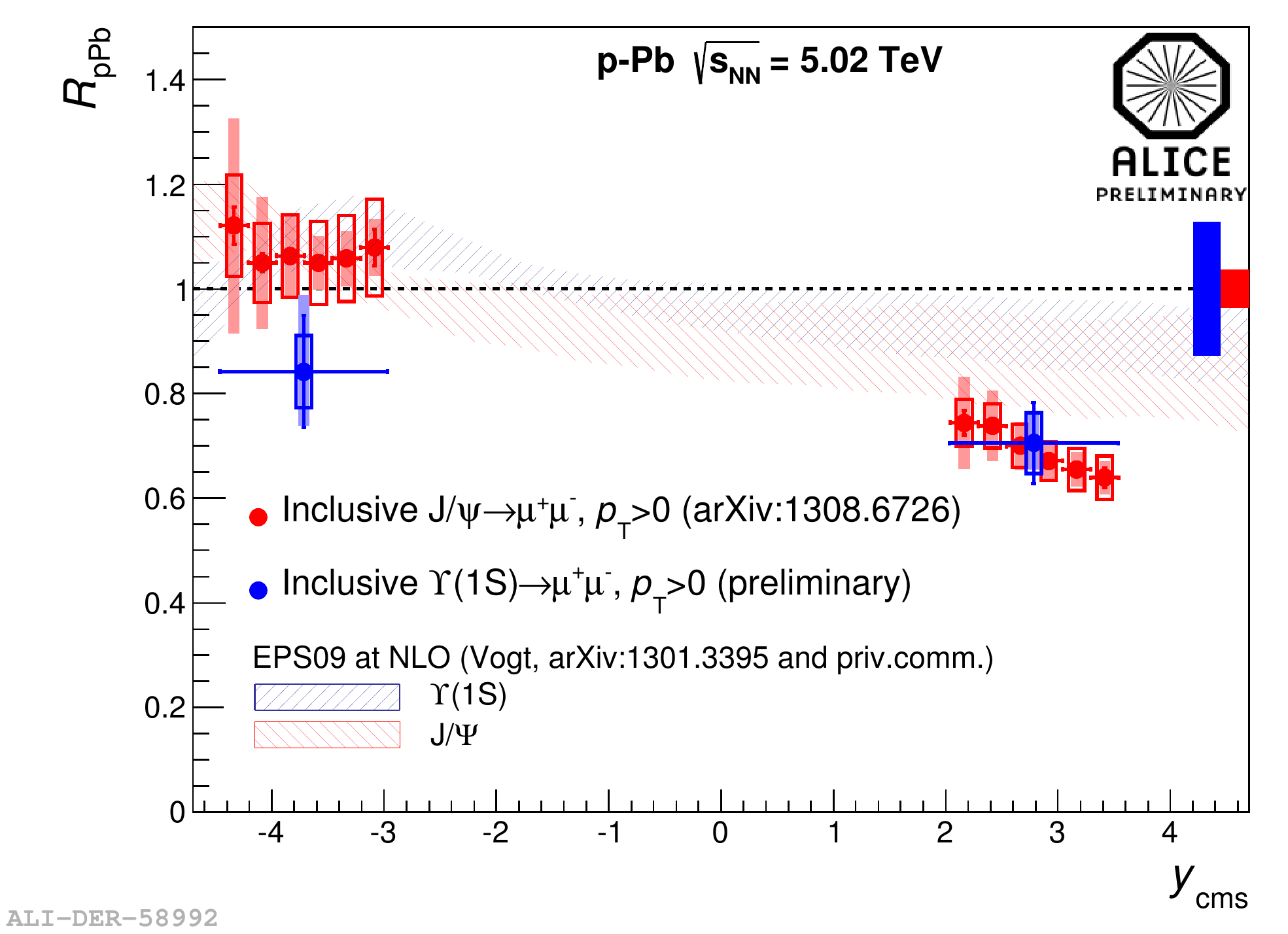}}
\resizebox{0.45\textwidth}{!}{
\includegraphics{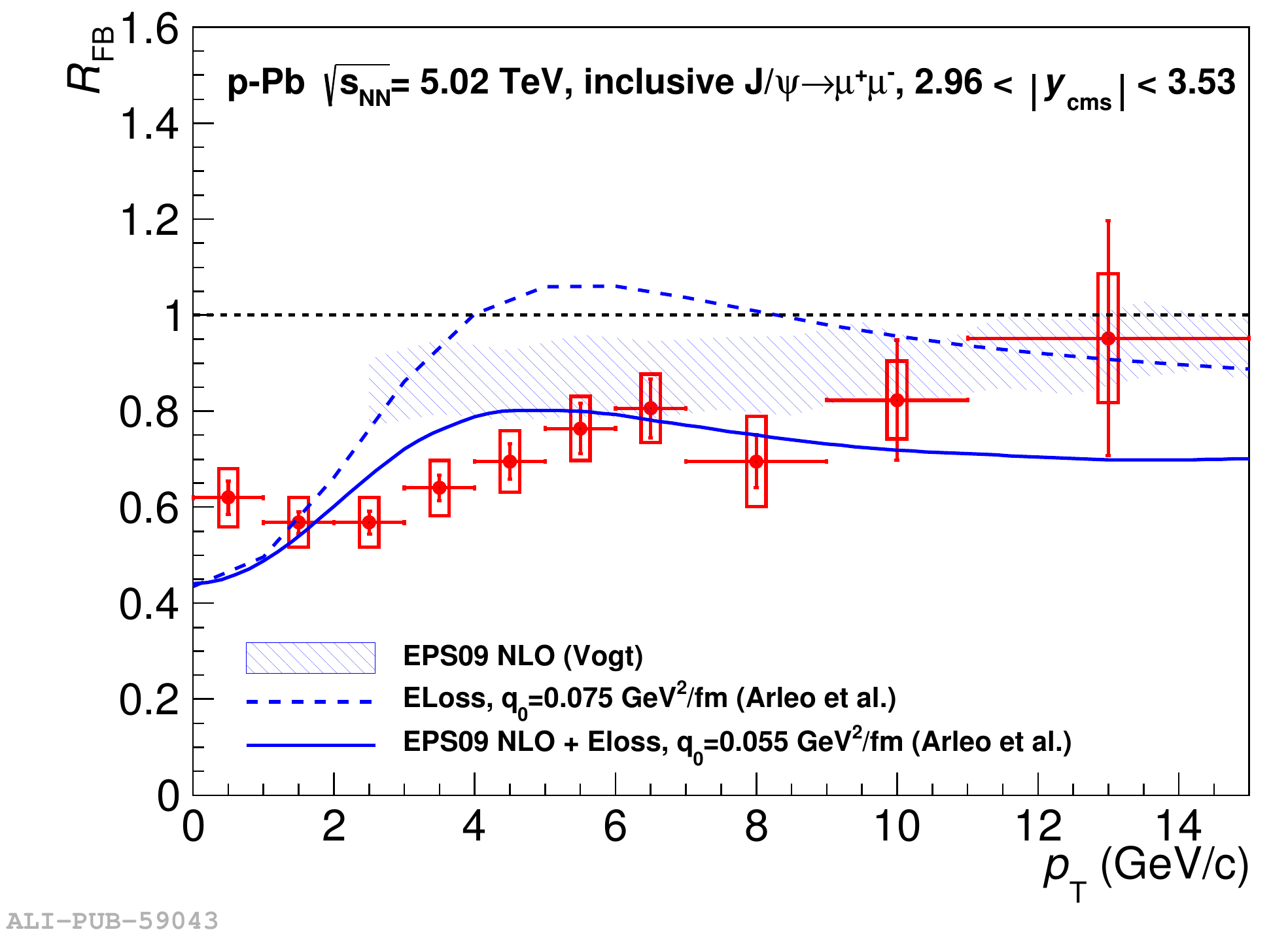}}
\caption{\label{Fig:R2}
Left panel: the nuclear modification factor for $\Upsilon({\rm 1S})$ and 
inclusive J/$\psi$   %production 
as a function of $y_{\rm cms}$ 
in p--Pb collisions at $\sqrt{s_{NN}} = 5.02$~TeV. 
Error bars correspond to statistical uncertainties, open boxes represent uncorrelated
systematic uncertainties, shaded boxes
show partially correlated systematic uncertainties and the boxes around $R_{\rm pPb} = 1$ shows the fully 
correlated uncertainties.  
Right panel: the forward to backward ratio for inclusive J/$\psi$  production as a function of $p_{\rm T}$. Statistical uncertainties are 
shown as bars, uncorrelated systematic uncertainties correspond to open boxes.
Theoretical model predictions are superimposed, see text for details.
}
\end{figure}

The nuclear modification factors for the J/$\psi$ and $\Upsilon({\rm 1S})$ as a function of the quarkonium rapidity 
%in the nucleon-nucleon center of mass rapidity ($y_{\rm CMS}$) 
$y_{\rm cms}$
are shown in the left panel of Fig.~\ref{Fig:R2}. The $R_{\rm FB}$ of J/$\psi$ as a function of $p_{\rm T}$  
is shown in the right hand panel of Fig.~\ref{Fig:R2}.  
The results are compared with theory predictions, based on a pure nuclear shadowing scenario, as
well as partonic energy loss, either in addition to shadowing or as the only nuclear effect 
(see~\cite{Palash,Igor} for a detailed discussion, and references therein). 
Within the  uncertainties, both the models based on shadowing only and the coherent energy loss 
approach are in fair agreement with  the data.   
The present theoretical and experimental uncertainties prevent from drawing more precise conclusions on the role
of the different contributions.  
%Instead, results from a  calculation 
The prediction 
based on the Color Glass Condensate model underestimates 
our results (not shown in Fig.~\ref{Fig:R2}, see~\cite{Igor} and references therein).  

\section{Conclusions}
ALICE has measured the production of quarkonia  in pp, p--Pb and Pb--Pb collisions at the energies of the LHC.    
Recent NLO NRQCD calculations that include color octet processes describe well the production in pp collisions. 
In Pb--Pb interactions, the results suggest that 
the 
%a %sizeable 
%fraction of the J/$\psi$ is produced by the  
(re-)combination of ${\rm c \bar{c}}$ pairs in the QGP or at the phase boundary 
plays a sizeable role in the J/$\psi$ production.  
The $\Upsilon({\rm 1S})$ has been found to be suppressed at forward rapidity to the same extent 
as measured at midrapidity by the CMS Collaboration.  
First p--Pb results are in fair agreement with predictions of models 
based on nuclear shadowing and coherent energy loss.

\end{document}